\documentclass[twocolumn,preprint,aps,10pt]{revtex4}
\usepackage{graphicx}% Include figure files
\def\beq{\begin{equation}}
\def\eeq{\end{equation}}
\def\bea{\begin{eqnarray}}
\def\eea{\end{eqnarray}}

\def \nn{\nonumber}

\begin{document}
\title{On the Complex Wilson Coefficients of New Physics Scenarios}
\author{L. Solmaz}
 \email{lsolmaz@balikesir.edu.tr}
\affiliation{Balikesir University, Physics Department.(10100),
Balikesir/Turkey}
 \author{S. Solmaz}
 \email{kerman@.metu.edu.tr}
\affiliation{Middle East Technical University, Physics Department. (06531),Ankara/Turkey}%

\date{\today}% It is always \today, today,
\begin{abstract}

%%%%%%%%%%%%%%%%%%%%%%%%%%%%%%%%%%%%%%%%%%%%%%%%%%%%%%%%%%%%%%%%%
We draw attention on the procedure, where  Standard Model
predictions and experimental results are compared and certain new
physics scenarios are ruled out, that requires great attention,
since there is still a room for new physics, especially when
contributions of extended models bear complex phases. We observe
that this is true even when SM and experiment yield  the same
results with zero uncertainties.

\end{abstract}
\maketitle

\section{Introduction and Conclusion}
%%%%%%%%%%%%%%%%%%%%%%%%%%%%%%%%%%%%%%%%%%%%%%%%%%%%%%%%%%%%%%%%%

At present for certain decays there are a number of impressive
theoretical predictions in great agreement with  the experimental
results. Due to  the consistency, uncertainties around one
$\sigma$ level seems acceptable. In general, the difference
between Standard Model (SM) results and experimental values are
seen on the half of New Physics (NP) scenarios and used to put
constraints on the new parameters(i.e. see Ref.\cite{Ref00}). If
SM predictions and experimental measurements are very close then
naturally NP is orphaned. The aim of the present work is to draw
attention on the side  effects of bounding Wilson coefficients
with the mentioned procedure.

 One of the best examples of such decays is the
 $B\rightarrow X_{s}\gamma$
decay, which has solid theoretical and experimental background,
widely used  to test and constrain a number of NP scenarios.
Recent review of this decay can be found in Ref.\cite{Ref01}.
%0310282}
 Taking into consideration NP effects, the present status of the
decay related with branching ratio can be expressed as

\bea\label{eq1} \eta\pm\sigma=\alpha \,|C|^2\,, \eea where
$\eta,\sigma$ denotes experimental measurements and uncertainties,
$\alpha$ satisfies proportionality related with the decay. With
the definition, $C=C^{SM}(\mu)+C^{NP}(\mu)$, $C^{SM,NP}(\mu)$ are
the Wilson coefficients at the low energy scale $\mu$.

 From the
experimental side, average of experimental measurements of the
branching ratio for
 the mentioned decay can be written as $Br(B\rightarrow
X_{s}\gamma)=(\eta\pm\sigma)\times10^{-4}$ with $\eta=3.34$,
$\sigma=0.38$ Refs.\cite{Ref02,Ref03,Ref04,Ref05,Ref06,Ref07}.
From the viewpoint of the SM the situation is very similar,
roughly $\eta\sim3.3-3.7$,6 $\sigma\sim0.3$
Refs.\cite{Ref08,Ref09,Ref10,Ref11,Ref12}.

 It can be dreamed that one day theoretical
and experimental results are very reliable and uncertainties are
under control, even $\sigma_{Theory,Experiment}\rightarrow 0$ is
approached. Meanwhile, $\sigma$ is commented as a nice room for a
number of new physics scenarios. Since, it can be used by NP, to
fill the gap between SM theory and experiment. In the case when
$\sigma$ is converging to zero and SM predictions are right in the
correct place, $C^{NP}\rightarrow 0$ and hence speculative
existence of new physics is ruled out.

However, what should be stressed is that, this  criteria is not
enough to reject the possibility of a hidden NP. To make it clear
assume the worst situation for NP, $C^{SM}$ satisfies the equality
given in Eq.\ref{eq1} without any theoretical or experimental
error, whence no room for NP scenarios,

\beq\label{eq2} \frac{\eta\pm 0}{\alpha}=
\,|C^{SM}(\mu)|^2=|C^{SM}(\mu)+ C^{NP}(\mu)|^2\,. \eeq

Notice that if $C^{NP}(\mu)$ is a real quantity, then $0$ is the
first solution for NP Wilson coefficient, second solution is $-2
C^{SM}(\mu)$. Once it is assumed complex in the form \beq
\label{eq3}C^{NP}(\mu)=C_r(\mu)+\it{i}C_i(\mu)\,,\eeq at first
sight ${C_r(\mu)=-C^{SM}(\mu),C_i(\mu)=\pm C^{SM}(\mu)}$ is the
solution, and the general solution for the NP scenario can be
written as

\beq\label{eq4} C_r^2 (\mu)+ C_i^2(\mu) = -2 C_r(\mu)
C^{SM}(\mu)\,, \eeq lying on the complex $C^{NP}(\mu)$ plane.
Complex phases in the final form of Wilson coefficients is
possible in certain extensions of the SM. As an example let us
mention one of the most popular extensions of the SM, Two Higgs
Doublet Model 2HDM(III) Ref.\cite{Ref13,Ref14,Ref15,Ref16}. In
this model, not only at the matching scale $\mu_W$ but also for
the $\mu_b$ scale there are complex couplings which can be written
in the appropriate form as follows

 \beq\label{eq5}
C_7^{eff}(\mu)=C_7^{SM}(\mu)+C_{7,r}^{2HDM}(\mu)+\it{i}
C_{7,i}^{2HDM}(\mu)\,.\eeq

\begin{figure}[htb]
\begin{center}
\vspace{0.5cm}
    \includegraphics[height=5cm,width=8cm]{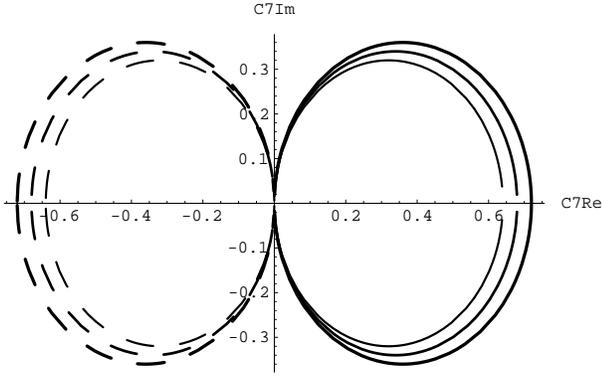}
    \vspace{0cm}
    \caption[]{Constraints on the new physics parameters $C_{7,r}^{NP},C_{7,i}^{NP}$ with the assumption
     $\eta_{Exp,SM}=3.5,\,\sigma_{Exp}=0.4.$ and  $\alpha\simeq 30.4\times10^{-4}$.
    Dashed lines (solid lines) show negative (positive) choices of $C_7^{SM}$, with increasing order of thickness
     with the $\sigma=\{-0.4, 0, 0.4\}$ choices respectively.}
    \label{fig1}
    \end{center}
\end{figure}
 Now, to see the effect of this approach, for simplicity, assume theoretical predictions of the SM and
 experimental results are given as follows
 \bea\label{eq6}
 Br(B\rightarrow X_{s}\gamma)_{SM}&=&(\eta_{SM}\pm 0)\times10^{-4}\,, \nn
 \\
 Br(B\rightarrow
 X_{s}\gamma)_{Exp}&=&(\eta_{Exp}\pm\sigma_{Exp})\times10^{-4}\,;
  \\
  &\,& \eta_{Exp,SM}=3.5,\,\,\sigma_{Exp}=0.4\,.\nn
 \eea
 In this approximation, Wilson coefficient of the SM takes $C_7^{SM}\sim\pm \{0.32,0.34,0.36\}$
 values in accordance with the error $\sigma_{Exp}$ is given in Eq.\ref{eq6}. and
 the proportionality constant  $\alpha\simeq 30.4\times10^{-4}$, is defined in Eq.\ref{eq1}.
 Notice that we assume no uncertainty for $\sigma_{SM}$ and set $\eta_{SM}=3.5$, hence neglect theoretical errors on purpose.
 Considering new physics, allowed ranges of the new Wilson
 coefficients $C_{7,r}^{NP},\,C_{7,i}^{NP}$
 can be extracted from Fig.\ref{fig1}. In the figure the choice
 $\sigma=0$ is presented in the middle of both left and right regions, stressing the possible
 solutions of $C_7^{NP}$ for $Br(B\rightarrow X_{s}\gamma)_{SM}=Br(B\rightarrow
 X_{s}\gamma)_{Experiment}$. By looking from one dimension if we
 set $C_{7,i}^{NP}=0$, possible values of $C_{7,r}^{NP}$ can be
 extracted from the same figure \ref{fig1}. As it can be deduced
 from the figure, considering complex phases enriches phenomenology.

 To summarize, scenarios permitting complex components in the final form of evolved Wilson coefficients
 have a rich potential for NP effects. Free parameters of NP should not be accepted as real
 from the beginning. Even if SM and experiment are in complete agreement, complex parts can help NP to
survive, when we consider the issue as an alternative solution.
While observing the possibility of such a structure of $New~
Physics$, branching ratio can not be used, solely, to refuse or
accept a new model. Nevertheless, since we have well motivated
theoretical and experimental background, it is possible to
back-transform Eq.\ref{eq4}. and obtain the most general form of
the NP scenarios at the matching scale. The price we have to pay
is, at least, two more unknowns which should be fixed by other
measurements or probably best by CP asymmetry $(A_{CP})$ of the
related decays.


\begin{thebibliography}{99}
\bibitem{Ref00} A. Ali, E. Lunghi, C. Greub, and G. Hiller, Phys. Rev. D 66
(2002) 034002.
%Improved Model-Independent Analysis of Semileptonic
%and Radiative Rare B Decays
\bibitem{Ref01}  T. Hurth, [arXiv:hep-ph/0212304].

\bibitem{Ref02}  R. Barate et al. [ALEPH Coll.], Phys. Lett. B 429 (1998) 169.

\bibitem{Ref03}  K. Abe et al. [Belle Coll.], Phys. Lett. B 511 (2001) 151.

\bibitem{Ref04} S. Chen et al. [CLEO Coll.], Phys. Rev. Lett. 87 (2001) 251807.

\bibitem{Ref05}  B. Aubert et al. [BABAR Coll.], hep-ex/0207074.

\bibitem{Ref06}  B. Aubert et al. [BaBar Coll.], hep-ex/0207076.

\bibitem{Ref07}  C. Jessop, SLAC-PUB-9610.

\bibitem{Ref08} A. Ali and C. Greub, Z. Phys. C49, 431 (1991);Phys. Lett. B 361, 146 (1995) [hep-ph/9506374]
% "Inclusive photon energy spectrum in rare B decays,"  "Photon
%energy spectrum in B › Xs. and comparison with data,"

\bibitem{Ref09} C. Greub, T. Hurth and D. Wyler, Phys. Lett. B 380, 385 (1996) [hep-ph/9602281];Phys. Rev. D 54, 3350 (1996) [hep-ph/9603404].
%"Virtual corrections to thedecay b › s.," "Virtual O(.s) corrections to the inclusive decay b› s.,"

\bibitem{Ref10} K. Adel and Y. Yao, Phys. Rev. D 49, 4945 (1994) [hep-ph/9308349].
%"Exact alphas calculation of b › s., b › sgluon,"

\bibitem{Ref11} K. Chetyrkin, M. Misiak and M. Munz, Phys. Lett. B 400, 206 (1997) [hep-ph/9612313].
%"Weak radiative B meson
%decay beyond leading logarithms,"

\bibitem{Ref12} K. Bieri, C. Greub and M. Steinhauser, Phys.Rev. D67 (2003) 114019

\bibitem{Ref13} D. Atwood, L. Reina and A. Soni, Phys. Rev. D53 (1996) 1199.

\bibitem{Ref14} D. Atwood, L. Reina and A. Soni, Phys. Rev. Lett. 75 (1995) 3800.

\bibitem{Ref15} D. Atwood, L. Reina and A. Soni, Phys. Rev. D 55 (1997) 3156.

\bibitem{Ref16} F. M. Borzumati and C. Greub , Phys. Rev. D 58 (1998)
    0784004.
 %BrzmtGrb9802391

\end{thebibliography}
\end{document}